\input{psfig.sty}
\documentclass[12pt,preprint]{aastex}

\begin{document}
 
\title{Disappearance of Hard X-ray Emission in the Last BeppoSAX Observation 
of the Z Source GX~349+2}

\author{R. Iaria\altaffilmark{1}, T. Di Salvo\altaffilmark{1,2}, 
N. R. Robba\altaffilmark{1},  
L. Burderi\altaffilmark{3}, L. Stella\altaffilmark{3}, 
F. Frontera\altaffilmark{4,5}, M. van der Klis\altaffilmark{2}}
\altaffiltext{1}{Dipartimento di Scienze Fisiche ed Astronomiche, 
Universit\`a di Palermo, via Archirafi n.36, 90123 Palermo, Italy;
iaria@gifco.fisica.unipa.it.}
\altaffiltext{2}{Astronomical Institute "Anton Pannekoek," University of 
Amsterdam and Center for High-Energy Astrophysics, Kruislaan 403, NL 1098 
SJ Amsterdam, the Netherlands; disalvo@science.uva.nl.}
\altaffiltext{3}{Osservatorio Astronomico di Roma, Via Frascati 33, 
00040 Monteporzio Catone (Roma), Italy}
\altaffiltext{4}{Physics Department, University of Ferrara, Via Paradiso 12, 
44100 Ferrara, Italy}
\altaffiltext{5}  {IASF - CNR Section of Bologna, Via P. Gobetti 101, 
40129 Bologna,  Italy}

\begin{abstract}

We report on the results from two BeppoSAX observations of the Z
source GX~349+2 performed in February 2001 and covering the broad
energy range 0.12--200~keV.  The light curve obtained from these
observations shows a large flaring activity, the count rate varying
from $\sim130$ to $\sim 260$ counts s$^{-1}$, indicating that the source
was in the flaring branch during these observations.  The average spectrum is
well described by a soft blackbody ($k T_{\rm BB} \sim 0.5$~keV) and a
Comptonized component having a seed-photon temperature of $k T_0 \sim
1$~keV, an electron temperature of $k T_{\rm e} \sim 2.7$~keV, and
optical depth $\tau \sim 11$.  To well fit the energy spectrum three
gaussian lines are needed at 1.2 keV, 2.6 keV, and 6.7 keV with 
corresponding equivalent widths of 13 eV, 10 eV, and 39 eV, probably 
associated to L-shell emission of Fe XXIV,  Ly$\alpha$ S XVI, 
and Fe XXV, respectively.
These lines may be produced at different distances from the neutron star, 
which increase when the count rate of the source increases. 
    An absorption edge is also needed at 9 keV with an
optical depth of $\sim 3 \times 10^{-2}$.  From the Color-Color Diagram (CD)
 we selected five zones from which we extracted the
corresponding energy spectra.  The temperatures of the blackbody and
of the Comptonized component tend to increase when the intensity of
the source increases.  We discuss our results comparing them to those
obtained from a previous BeppoSAX observation, performed in March
2000, during which the source was a similar position of its Z-track.
In particular we find that, although the source showed similar spectral 
states in the 2000 and the 2001 observations, a hard tail, that was 
significantly detected in March 2000, is not observed in these recent 
observations.
\end{abstract}


\section{Introduction}

Low Mass X-ray Binaries (LMXB) containing old and low magnetic-field neutron 
stars (NS) are usually divided into Z and Atoll sources, according to the 
path they describe in an X-ray Color-Color Diagram (CD) or 
hardness-intensity diagram (Hasinger \& van der Klis 1989) assembled by 
using the source count rate over a typical (usually 2--20 keV) X-ray energy 
range. Atoll sources are usually characterized by relatively low 
luminosities ($\sim 0.01-0.2\ L_{\rm Edd}$) and some show transient 
behavior, while the six known Z sources in the Galaxy are among the most 
luminous LMXBs, accreting persistently close to the Eddington limit 
($L_{\rm Edd}$) for a 1.4 $M_\odot$ NS.  The instantaneous position of an
individual source in the CD, which determines most of the observed spectral 
and temporal properties of the source, is thought to be an indicator of the 
mass accretion rate (e.g.\ Hasinger et al. 1990; see van der Klis 1995 for 
a review). It has been suggested that the mass accretion rate 
(but not necessarily the X-ray luminosity) of individual sources 
increases along the track from the top left to the bottom right, i.e.\ 
from the islands to the banana branch in atoll sources and from the 
horizontal branch (hereafter HB) to the normal branch (NB) and to the
flaring branch (FB) in Z sources.

However, while there is a general correlation between temporal variability
properties (in particular frequencies of the observed quasi periodic 
oscillations, QPOs, see van der Klis 2000 for a review) and position
in the CD, the correlation with the source X-ray count rate or X-ray flux in 
the 2--50~keV energy range is complex; a correlation is observed on short 
time scales (hours to days), but not on longer time scales. 
A possible explanation of this behavior is that most 
timing and spectral parameters are determined by the dynamical properties of 
the accretion disk, while the X-ray flux is determined by the total accretion
rate, which may be different from the instantaneous accretion rate through 
the disk if, for instance, matter can flow radially close to the NS
(see e.g.\ van der Klis 2001). 

Hard X-ray spectra extending up to energies of several tens to hundred 
keV have been revealed in about 20 ``faint'' NS LMXBs (some of them are
confirmed Atoll sources, see Di Salvo \& Stella 2002 for a review).
In these systems a power law-like component is observed, with typical photon 
indices of $\sim 1.5 -2.5$, and a high energy exponential cutoff between 
$\sim 20$ and many tens of keV.
This component is interpreted in terms of unsaturated thermal Comptonization. 
Sometimes, in the so called "hard state" of atoll sources, there is no 
evidence for a cutoff up to $\sim 100-200$~keV. 
Some sources appear to spend most of the time 
in this state (e.g.\ 4U 0614+091, Ford et al. 1996; Piraino et al. 1999, and
references therein). In others a gradual transition from the soft to the 
hard state has been observed in response to a decrease of the source 
X-ray luminosity and/or the source drifting from the banana branch 
to the island state. This transition is often modelled in terms of a gradual 
decrease of the electron temperature (and increase of the optical depth) of 
the Comptonizing region. 

On the other hand, the spectrum of the Z sources is much softer, with
cutoff energies usually well below 10~keV.  However, hard tails were
occasionally detected in their spectra. 
A variable hard component dominating the spectrum of Sco X--1 above 
$\sim 40$~keV was detected as early as 1966 (Peterson \& Jacobsen 1966; 
see also Riegler et al. 1970; Agrawal et al. 1971; Haymes et al. 1972). 
In other occasions the hard tail in Sco X--1 was not found (e.g., Miyamoto 
\& Matsuoka 1977, and references therein; Soong \& Rothschild 1983; 
Jain et al. 1984; Ubertini et al. 1992), perhaps owing to pronounced 
variations. 
Evidence for a hard component was also found in Cyg~X--2 (Peterson 1973),
GX~349+2 (Greenhill et al. 1979), and in Ginga data of GX 5--1 (although
in this case a contribution from a contaminating source could not be 
excluded, Asai et al. 1994). 

Renewed interest in the hard X-ray emission properties of luminous 
LMXBs was motivated by some recent observations, mostly using the broad band
capabilities of BeppoSAX (0.1--200 keV) and RXTE (2--220 keV). 
Recently a hard tail was detected in GX~17+2, observed by BeppoSAX. In 
this case the intensity variations of the hard tail were clearly correlated 
with the source spectral state: a factor of 20 decrease was observed moving
from the HB to the NB (Di Salvo et
al. 2000).  The presence of a variable hard tail in Sco X--1 was
recently confirmed by OSSE and RXTE observations (Strickman \& Barret
2000; D'Amico et al. 2001). A hard tail was also detected in GX 349+2
(Di Salvo et al. 2001, hereafter Paper I) and Cyg~X--2 (Frontera et al. 
1998; Di Salvo et al. 2002), as well as in the peculiar bright LMXB Cir X--1 
(Iaria et al. 2001). These hard components can be fitted by a power law, 
with photon index in the range 1.9--3.3, contributing up to 10\% of the 
total source luminosity. 

GX~349+2, also known as Sco X--2, was called an odd-ball among the Z 
sources (Kuulkers \& van der Klis 1998).
Similar to the case of Sco X--1, GX~349+2 shows a short and underdeveloped 
HB (if at all).  The source variability in the frequency range below 
100 Hz is closely correlated with the source position on the X-ray CD, 
as in other Z sources.  Quasi periodic oscillations at kHz frequencies 
(kHz QPO) were detected in the NB of its Z-track (Zhang, Strohmayer, \& 
Swank 1998).  However, GX~349+2, which sometimes shows broad 
noise components changing not only with the position in the Z, but 
also as a function of the position in the hardness-intensity diagram, 
differs somewhat from the other Z sources and shows similarities to the 
behavior seen in bright atoll sources, such as GX 13+1 and GX 3+1 
(Kuulkers \& van der Klis 1998; see also O'Neill et al. 2001, 2002).   
Using a BeppoSAX observation performed in March 2000, Di Salvo et al. 
(2001) showed that the source energy spectrum below 30 keV could be well 
fit by a blackbody (with a temperature of 0.5--0.6 keV) and a Comptonized 
component (with seed-photon temperature of $\sim 1$ keV and electron 
temperature of $\sim 2.7$ keV). 
Three discrete features were observed in the spectrum:
an emission line at 1.2 keV, probably associated to Ne X or a complex
L-shell of Fe XXIV, an emission line at 6.7 keV and an absorption edge 
at 8.5 keV, both corresponding to emission from the K-shell of 
highly-ionized iron (Fe XXV).  
Above 30 keV, during the non-flaring state, a power-law component was 
significantly detected having a photon index of $\sim 1.9$ and a flux of 
$\sim 1.2 \times 10^{-10}$ erg cm$^{-2}$ s$^{-1}$ in the energy band 
10--60 keV. 
Because there was no evidence of contaminating sources, the authors 
concluded that the most probable candidate for the hard emission was 
GX 349+2 itself.
The hard component detected in GX~349+2 is one of the hardest among the 
high energy components detected so far in bright LMXBs, with no evidence 
for a high energy cutoff in the BeppoSAX range (up to $\sim 100$~keV).

In this paper we report the results of a spectral study of a long
($\sim 200$ ks) BeppoSAX observation of GX~349+2 in the energy range
0.1--200~keV.  We find that, although the spectrum below 30 keV is
similar to that observed during the observation reported on Paper I,
the hard component is not present (or significantly weaker) during our 
observation, demonstrating that these hard tails can be highly variable
and are probably not univocally related to the position in the CD.

\section{Observations}

The Narrow Field Instruments (NFI) on board the BeppoSAX satellite are
four co-aligned instruments which cover more than three decades in
energy, from 0.1~keV up to 200~keV, with good spectral resolution over
the whole range (see Boella et al. 1997a for detailed description of
BeppoSAX instruments).  These are two Medium Energy Concentrator
Spectrometers, MECS (position sensitive proportional counters
operating in the 1.3--10~keV band, Boella et al. 1997b), a Low Energy
Concentrator Spectrometer, LECS (a thin window position sensitive
proportional counter with extended low energy response, 0.1--10~keV;
Parmar et al. 1997), a High Pressure Gas Scintillation Proportional
Counter (HPGSPC; energy range of 7--60 keV; Manzo et al. 1997) and a
Phoswich Detection System (PDS; energy range of 13--200 keV; Frontera
et al. 1997).

In this paper we report on two BeppoSAX observations of GX~349+2. The
source was observed from 2001 February 12 11:23 UT to 2001 February 13
17:43 UT and from 2001 February 17 16:15 UT to 2001 February 19 20:42
UT. The total effective exposure times are $\sim 30$ ks for the LECS, $\sim
138$ ks for the MECS, $\sim 130$ ks for the HPGSPC, and $\sim 66$ ks
for the PDS.  We selected the data for the spectral analysis in
circular regions centered on the source with $8'$ and $4'$ radius for
LECS and MECS, respectively. The background subtraction was obtained
with standard methods by using blank sky observations.  The background
subtraction for the high-energy (non-imaging) instruments was obtained
by using off-source data for the PDS and Earth occultation data for
the HPGSPC.  

In Figures \ref{fig1} and \ref{fig2} (lower panels) we show the 200 s
binned MECS light curve of GX~349+2 in the 1.8--10.5~keV range for the
first and the second observation, respectively; the light curve presents 
a large variability and the count rate varies from 130 c/s up to 260 c/s.  
In the upper and middle panels we plot the Soft Color (SC), i.e.  the ratio
of the counts in the 4.5--7~keV to the 1.8--4.5~keV band, and the
Hard Color (HC), i.e. the ratio of the counts in the 7--10.5~keV to 
the 4.5--7~keV energy band, as functions of time.

In Figure \ref{fig3zero} we show the CD of GX~349+2 , where the HC and
the SC are as defined above. The red triangles and the black stars
indicate the positions of the source in the CD during a previous BeppoSAX
observation taken in March 2000 (see Paper I) and during our observations, 
respectively. From this figure it is clear that the position in the Z-track
was approximately the same during all the observations. To study 
the spectral variability along the CD we selected five regions 
in the CD: SC$<$ 0.47 (interval 1), 0.47 $<$SC$<$ 0.50 (interval 2),
0.50 $<$SC$<$ 0.55 (interval 3), 0.55 $<$SC$<$ 0.60 (interval 4),
SC$>$ 0.64 (interval 5), from which we extracted the corresponding
count spectra.

In Figure \ref{fig3bis} we show the Hardness-Intensity Diagram (HID)
where the Color on the y-axis is the SC and the Intensity is the
count rate of the source in the energy band 1.8--10.5 keV. The SC is
a linear function of the Intensity. We fitted the points using a
linear relation $y = mx +q$ obtaining as best fit $m = (1.65 \pm
0.02) \times 10^{-3}$ and $q = 0.235 \pm 0.004$. This allows us to
associate to the regions selected in the CD the corresponding count
rate.  In Table \ref{tab:TABexpo} we report the exposure times of the
four instruments for each selected region and the corresponding count
rate.

\section{Spectral Analysis} 

We extracted an averaged spectrum from the whole observation which is
discussed in \S 3.1. In section \S 3.2 we present the results obtained
from the spectra selected in different regions of the CD.
Relative normalizations of the four NFIs were treated as free
parameters in model fitting, except for the MECS normalization
which was fixed to a reference value of 1. We checked after the fitting
procedure that these normalizations were in the standard range for
each instrument\footnote{See the BeppoSAX handbook at
http://www.sdc.asi.it/software/index.html.}.  The energy ranges used
in the spectral analysis were: 0.12--3.5~keV for the LECS, 1.8--10~keV
for the MECS, 7--30~keV for the HPGSPC, and 15--200~keV for the PDS.
We rebinned the energy spectra in order to have approximately the same
number of bins per instrument resolution element across the entire
energy range. 
A 1\% systematic error was applied to the  spectra selected in different 
regions of the CD, and a 2\% systematic error was applied to the averaged  
spectrum to take into account possible variability of the spectrum during
the observation.

\subsection{The Averaged Spectrum}

We extracted an averaged spectrum from the two BeppoSAX observations. 
Following the results of Paper I, we started fitting the continuum with 
a blackbody plus a Comptonized component ({\tt Comptt}, Titarchuk 1994)
obtaining a $\chi^2$(d.o.f.) of 309(197).   In the
residuals with respect to this model (see Fig.\ref{fig4}, middle
panel) we find the presence of a weak hard excess at energies higher
than 35 keV. Evident localized excesses are also present around 1.2 keV, 
2.6 keV and 6.7 keV and an absorption feature is present between 9 and 
10 keV.  

The fit can be significantly improved by the addition of the components
described below. We added a power-law component with photon index of $ 1.52$
to fit the hard excess; an F-test gives a probability of chance improvement 
of the fit for the addition of this component of $ 1.9 \times 10^{-7}$.  
We added an absorption edge to fit the feature between 9 and 10 keV; 
the energy of the absorption edge is $\sim 9$ keV and its optical depth 
is $\tau \sim 3 \times 10^{-2}$.
To fit the localized excesses we used three narrow gaussian emission lines.
Initially we added  a gaussian line having its centroid at $ 1.18$ keV and 
equivalent width $\sim 13$ eV, its addition  gave  a probability of chance 
improvement of $1.1 \times 10^{-8}$; then we added a gaussian line having 
its centroid at $ 2.6$ keV and equivalent width $\sim 10$ eV, its addition  
gave a probability of chance improvement of $\sim 2.1 \times 10^{-5}$; and 
finally we added a gaussian line having its centroid at $ 6.75$ keV and  
equivalent width $\sim 39$ eV, its addition gave  a probability of chance 
improvement of $\sim 1.6 \times 10^{-9}$.  
The $\chi^2$(d.o.f.) of the best fit model is  149(184). 

In Figure \ref{fig4} (upper and lower panels) we plot the data and the
residuals corresponding to the best fit model described above and in
Figure \ref{fig6} we plot the corresponding unfolded spectrum. In
Tables \ref{tab:TABavera} and \ref{tab:TABaverabis} we report the
parameters of the continuum and of the narrow features, respectively.
The flux of the blackbody and of the Comptonized component in the
energy band 0.1-200 keV are $ 0.45 \times 10^{-8} $ erg cm$^{-2}$
s$^{-1}$ and $\sim 2 \times 10^{-8} $ erg cm$^{-2}$ s$^{-1}$,
respectively.  The flux of the power-law component in the energy band
10-60 keV is $\sim 2.2 \times 10^{-11} $ erg cm$^{-2}$ s$^{-1}$,
approximately one order of magnitude lower than the hard power-law
flux during the observation reported on Paper I.

\subsection{Spectral analysis as a function of the source position in the CD}

We fitted the five spectra extracted from different positions in the
source CD (see Fig. \ref{fig3zero}) using the model adopted for the
averaged spectrum, keeping the photon index of the power-law component
fixed at 1.52, the value obtained from the fit of the averaged
spectrum.  This model gives an acceptable fit for all the spectra.  In
Tables \ref{tab:TAB2} and \ref{tab:TAB2bis} we report the parameters
of the continuum and of the narrow features for each spectrum. Going
from interval 1 to interval 5, that is going to higher intensity of
the source, the flux of the blackbody component increases from $
0.39 \times 10^{-8} $ erg cm$^{-2}$ s$^{-1}$ to $ 0.50
\times 10^{-8} $ erg cm$^{-2}$ s$^{-1}$, the flux of the
Comptonized component increases from $\sim 1.45 \times 10^{-8} $ erg
cm$^{-2}$ s$^{-1}$ to $\sim 3.0 \times 10^{-8} $ erg cm$^{-2}$
s$^{-1}$; the temperatures of the blackbody and of the Comptonized
components also increase. 
In Figure \ref{trend} we show the fluxes and the temperatures of the 
blackbody and Comptonized components as function of the count rate.  

The addition of the gaussian line at $ 1.18$ keV is statistically 
significant in intervals 1, 2 and 3, and its equivalent width varies 
between 12 and 20 eV.  The addition of the gaussian line at $ 2.6$ 
keV is  significant in intervals 1, 2, 3 and 4 and its equivalent width 
is $8$ eV.  The addition of the the gaussian line at $ 6.7$ keV 
is statistically significant in each interval and its equivalent width 
decreases from 80 eV to 18 eV going from interval 1 to interval 5. 
The decrease of the equivalent width is due to both an increase of the
continuum flux and a decrease of the line intensity. 
The addition of an absorption edge at $\sim 9$ keV, with an optical depth 
of $\sim 0.03$, is always statistically significant.  The addition of the 
power-law component is significant in interval 1 and interval 3 and the 
flux, in the energy band 10--60 keV is between $ 1.4 \times 10^{-11}$ 
erg cm$^{-2}$ s$^{-1}$ and $ 2.8 \times 10^{-11} $ erg cm$^{-2}$ 
s$^{-1}$, as reported in Table \ref{tab:TAB2}, similar to the value found 
for the averaged spectrum.

\section{Discussion}

We studied the BeppoSAX energy spectra of GX~349+2, extracted at
different positions of the source in the CD. The best fit model up to
energies of $\sim 30$~keV consists of a blackbody and a Comptonization
spectrum (described by the {\tt Comptt} model), three emission lines
and an absorption edge.

The equivalent hydrogen column, N$_H$, derived from the best-fit model
is $\sim 0.67 \times 10^{22}$ cm$^{-2}$, in agreement with the results
of Paper I. 
Cooke \& Ponman (1991), using data from the medium and low energy
X-ray detectors on EXOSAT, obtained a value of $(0.86 \pm 0.01) \times
10^{22}$ cm$^{-2}$, and evaluated the visual extinction in the direction 
of the source to be $A_V = 3.9 \pm 0.5$ mag; a similar value of 
$\sim 0.8 \times 10^{22}$ cm$^{-2}$, which implies a distance to GX 349+2 
of 5 kpc, was reported by Christian \& Swank (1997) using the Einstein 
solid-state spectrometer (SSS; 0.5--4.5 keV).
Using the value of the equivalent hydrogen column to the source that we 
obtain from the BeppoSAX observations, we can recalculate the visual
extinction from the observed correlation between visual extinction and 
absorption column (Predehl \& Schmitt, 1995); this gives 
$A_V \simeq 3.6-4.1$ mag, still compatible with the values previously reported 
(Cooke \& Ponman 1991; Penninx \& Augusteijn 1991). In the direction of 
GX 349+2, this would correspond to a distance between 3.6 and 4.4 kpc 
(Hakkila et al. 1997).
These values are not very different from 5 kpc. Therefore we will continue
to adopt 5 kpc for the distance to GX 349+2 to facilitate the comparison
with previous results. Note, however, that all the luminosities quoted here
can be lower by a factor of $\sim 1.3-1.9$.

The blackbody temperature is $kT_{\rm BB} \sim 0.5-0.6$~keV, its
luminosity, in the energy range 0.1--200 keV, is $\sim 1.34 \times
10^{37}$ erg s$^{-1}$ and the radius of the blackbody emitting region
is $R_{\rm BB} \sim 36$~km. The blackbody temperature and flux both 
increase with increasing the source luminosity (see Fig.~\ref{trend}), 
while the blackbody radius does not significantly change.
The luminosity of the Comptonized component, in the 0.1-200 keV band,
is $\sim 3.4 \times 10^{37}$ erg s$^{-1}$, 80\% of the total
luminosity, and is $\sim 20\%$ of the Eddington luminosity for a $1.4\;
M_\odot$ NS.  The temperature of the soft seed photons for the 
Comptonization is $k T_{\rm 0} \sim 1$~keV.  These are Comptonized in a 
hotter ($k T_{\rm e} \sim 3$~keV) region of moderate optical depth 
($\tau = 10-12$ for a spherical geometry).  
The radius of the region emitting the seed-photon Wien
spectrum, calculated as in In 't Zand et al. (1999), is $R_{\rm W} = 3
\times 10^4 D \sqrt{{\rm Flux}_{comptt}/(1+y)}/(kT_0)^2\; {\rm km}= 7-9$~km, 
where $D$ is the distance in kpc, Flux$_{comptt}$ is the bolometric flux 
of the Comptonization component in ergs cm$^{-2}$ s$^{-1}$, $y$ is the 
Compton parameter and $k T_0$ is in keV. 
We note that $R_{\rm W}$ is comparable with the  radius of a NS. 
The seed photon temperature significantly increases with the source 
luminosity (see Fig.~\ref{trend}). There is small evidence of an increase 
of the optical depth and a decrease of the radius of the seed photons 
emitting region with increasing the source luminosity, while the electron 
temperature is consistent with being constant.

A broad ($\sim 0.7$~keV FWHM) iron K-shell emission line is present
at $\sim 6.7$~keV, with equivalent width between 18 and 85 eV, which
decreases with increasing the source luminosity. This is accompanied by 
an absorption edge at $\sim 9$~keV with optical depth $\tau \sim 0.04$. 
The high energy of both the iron line and edge indicates that these 
features are produced in a highly ionized region (corresponding 
approximately to Fe XXV).  
We also found evidence for an emission line at $\sim 1.2$~keV, with
equivalent width of $ 9-20$ eV slightly decreasing with increasing 
the source luminosity, which can be associated with emission from the 
L-shell of Fe~XXIV (see e.g. Kallman et al. 1996), 
and an emission line at $ 2.6$~keV, with equivalent width of 
$ 10$ eV, which can be associated with Ly$\alpha$ emission from S XVI.

The ionization parameters, Log$_{10} \xi$, corresponding to S XVI, Fe
XXIV and Fe XXV are 3.1, 3.04 and 3.67, respectively. From $\xi = L_x/
n_e r^2$ (see Krolik, McKee \& Tarter 1981) and $L_{\rm line}= 4 \pi D^2
I_{\rm line} = n_e^2 V \alpha A f$ we can obtain the distance, $r$, from the
central source, and the corresponding densities, of the region where the
lines are produced.  
In these formulas $L_x$ is the total unabsorbed luminosity
of the source, $n_e$ the electron density, $L_{\rm line}$ the luminosity
of the line, $D$ the distance to the source, $I_{\rm line}$ the intensity
of the line, $V$ the emitting volume, $\alpha$ the recombination
parameter, $A$ the cosmic abundance of the element and $f$ the
fractional number of ions in the given ionization state of 
the considered element. 
We assumed a spherical volume of radius $r$ and fixed $f=1$. The
recombination parameter $\alpha$ was obtained using the relation and the
best fit parameters for S XVI, Fe XXIV and Fe XXV reported by Verner \&
Ferland (1996), where we fixed the plasma temperature at the electron
temperature of the Comptonizing cloud.  For the averaged spectrum we
find that the Fe XXV line is produced at $r = 5.1 \times 10^9$ cm
with a corresponding electron density of $n_e = 5.9 \times 10^{14}$
cm$^{-3}$, the S XVI line at $r = 1.0 \times 10^{10}$ cm with $n_e
= 5.7 \times 10^{14}$ cm$^{-3}$ and, finally, the Fe XXIV line at
$r = 2.6 \times 10^{10}$ cm with $n_e = 9.2 \times 10^{13}$
cm$^{-3}$.  
On the other hand if we assume that the Comptonizing cloud
is indeed an Accretion Disc Corona (ADC), where the coronal
temperature is the electron temperature of the cloud,  then the coronal
radius is $R_c \sim M_{\rm NS}/M_{\odot} \; T_7^{-1}$ R$_{\odot}$ (White
\& Holt 1982), where $M_{\rm NS}$ is the NS mass and $ T_7$ the
coronal temperature in units of $10^7$ K. Assuming a
NS mass of 1.4 $M_{\odot}$ and the electron temperature
obtained from the fit we find $R_c \sim 3
\times 10^{10}$ cm. This implies that the  Fe XXV line is
produced at $r \sim 0.17 R_c$, the S XVI line at $r \sim 0.34 R_c$ and
the Fe XXIV line at $r \sim 0.87 R_c$.  The Fe XXV line is broader
than the S XVI and Fe XXIV lines probably because it is produced in
the inner region of the corona and the the escaping photons are more
affected by Compton scattering.  

A weak narrow absorption feature
is present at $\sim 4.2$ keV.  We fitted this component using an
absorption edge and found an upper limit of the optical depth of
$\tau \sim 3.82 \times 10^{-2}$. This absorption edge could correspond
to K-edge of Ar XVII, dominating at a ionization parameter of 
Log$_{10} \xi \sim 3$. The corresponding Ar XVII emission line should be 
found at $\sim 3.14$ keV. Supposing that it is produced in the same
region of Fe XXIV, that is the external region of the corona, and using the 
parameters of density and radius obtained for Fe XXIV, the cosmic abundance 
of Ar and the corresponding recombination parameter (Verner \& Ferland
1996) we obtain an intensity of the Ar XVII line of $\sim 1.1 \times
10^{-3}$ photons cm$^{-2}$ s $^{-1}$; from the fit we find a
compatible upper limit of the intensity of $\sim 2 \times 10^{-3}$
photons cm$^{-2}$ s $^{-1}$ implying that the Ar XVII line may be
overwhelmed by the continuum emission.  

We repeated the same procedure
for the spectra extracted in the different region of the CD in order
to study the emission region of the lines at different source luminosities;
the results, for the statistically significant lines, are reported in
Table \ref{tab:line}.  It is evident that when the count rate
increases the formation region of the lines moves at larger radii
where the density is lower.

One of the main results of this paper regards the hard component.
At energies higher 30 keV a hard excess is indeed present in our 
observations. However, the 10--60 keV flux of the power-law component
that we used to fit this hard excess is one order of magnitude lower
than that measured during the BeppoSAX observation of March 2000 (Paper I),
although the source was in a similar position of the Z track.  In Paper I a
hard component was required to match the spectrum above 30~keV during
the non-flaring emission.  This component could be fit by a power-law
with photon index $\sim 2$ and with a flux of $1.2 \times 10^{-10}$
ergs cm$^{-2}$ s$^{-1}$ in the energy range 10--60 keV.  In our
observations the photon index of the power law is 1.5 (still compatible 
with the previous value), while the flux in the same energy band is 
$\sim 2 \times 10^{-11}$ ergs cm$^{-2}$ s$^{-1}$. 
Indeed we cannot exclude that the power law we fit during our
observations is due to the hard diffuse emission of the Galactic ridge. 
Using data from Valinia \& Marshall (1998), for latitudes
$1.5^{\circ}<|b|<4^{\circ}$ and longitudes $|l|<15^{\circ}$ (the
region of GX~349+2) the flux of the diffuse Galactic emission is $\sim
3.2 \times 10^{-11}$ ergs cm$^{-2}$ s$^{-1}$ in the 10--60 keV energy
range for the effective solid angle of the PDS FOV and the photon
index of the power-law is $1.7 \pm 0.2$.  These values are compatible
to those we found in our analysis of GX 349+2.
We conclude that the hard power-law component during these new BeppoSAX
observations is much weaker or even absent.

A hard power-law component was observed in several Z sources, indicating 
that this is probably a common feature of these sources.  The presence (or
strength) of these components appears sometimes to be related to the source
state or its position in the CD. The only clear example of this behavior
was given by a BeppoSAX observation of GX 17+2. In this source the intensity 
of the hard component (a power-law with photon index of $\sim 2.7$) 
showed the strongest intensity in the HB of its CD; a factor of $\sim 20$ 
decrease was observed when the source moved from the HB to the NB, i.e.\ 
from low to high inferred mass accretion rate.  For other sources some
evidence was found that the hard component becomes weaker for higher 
accretion rates (GX~5--1, Asai et al. 1994; GX~349+2, Paper I; 
Cyg X--2, Di Salvo et al. 2002; Cir X--1, Iaria et al. 2001). However,
in recent RXTE/HEXTE observations of Sco X--1, a hard power-law tail was 
detected in 5 out of 16 observations, without any clear correlation with the 
position in the CD (D'Amico et al. 2001). GX 349+2 may show a behavior that
is similar to the one observed in Sco X--1. Indeed GX 349+2, as well as 
GX 17+2, is classified as a Sco-like source, which are thought to have a 
lower inclination than the other Z sources (referred to as Cyg-like sources; 
Kuulkers et al. 1994; Kuulkers \& van der Klis 1995), and similarly to 
Sco X--1 (but not to GX 17+2), does spend a relatively short time in the HB 
(i.e.\ at the lowest inferred mass accretion rates). 

The behavior of Sco X--1 and GX 349+2 suggests that there 
might be a second parameter, besides mass accretion rate or position of the
source in the Z-track, regulating the presence of hard emission in these 
systems. As already mentioned, the second parameter regulating the spectral 
state transitions might be the truncation radius of the optically thick 
disc. However, what determines the radius at which the disc is truncated 
is not clear yet: this could be the mass accretion rate through the disk 
normalized by its own long-term average (as proposed by van der Klis 2001), 
but also magnetic fields, the fraction of power dissipated in a hot, 
optically thin, corona (see e.g.\ Chen 1995), or the formation/quenching 
of a jet could play a role.
Interestingly, Strickman \& Barret (2000) suggest that the hard X-ray 
emission present in Sco X--1 data from OSSE may be correlated with periods 
of radio flaring.

This might be generally true for this kind of systems.
Although the mass accretion rate appears to be the main parameter 
driving the spectral hardness of atoll sources, 
there is evidence that at least on occasions an additional parameter 
controls the soft/hard spectral transitions. A clear example is given by 
a recent observation of 4U 1705--44, in which the source underwent a soft to 
hard state transition while the 0.1--200~keV bolometric luminosity of the 
source decreased by a factor of $\sim 3$ from the soft to the hard state 
and increased by only a factor of $\sim 1.2$ in the opposite transition from 
the hard to the soft state (Barret \& Olive 2002).
On another occasion the same source displayed hard and soft states in which 
the source luminosity was different by a much larger factor, up to one order 
of magnitude. Again, the second parameter regulating the spectral state 
transitions might be the truncation radius of the optically thick disc.  
Note also that even though variations of the mass accretion rate appear to be 
the main cause of the spectral transitions of accreting black hole candidates 
(BHCs), there is some evidence that a second, yet unknown, parameter can give 
rise to these transitions. The existence of a second parameter was indeed 
proposed to explain the soft/hard spectral transitions observed in the BHCs 
XTE~J1550--564 (Homan et al. 2001) and GS~2000+25 (Tanaka 1989).

\begin{figure}
\plotone{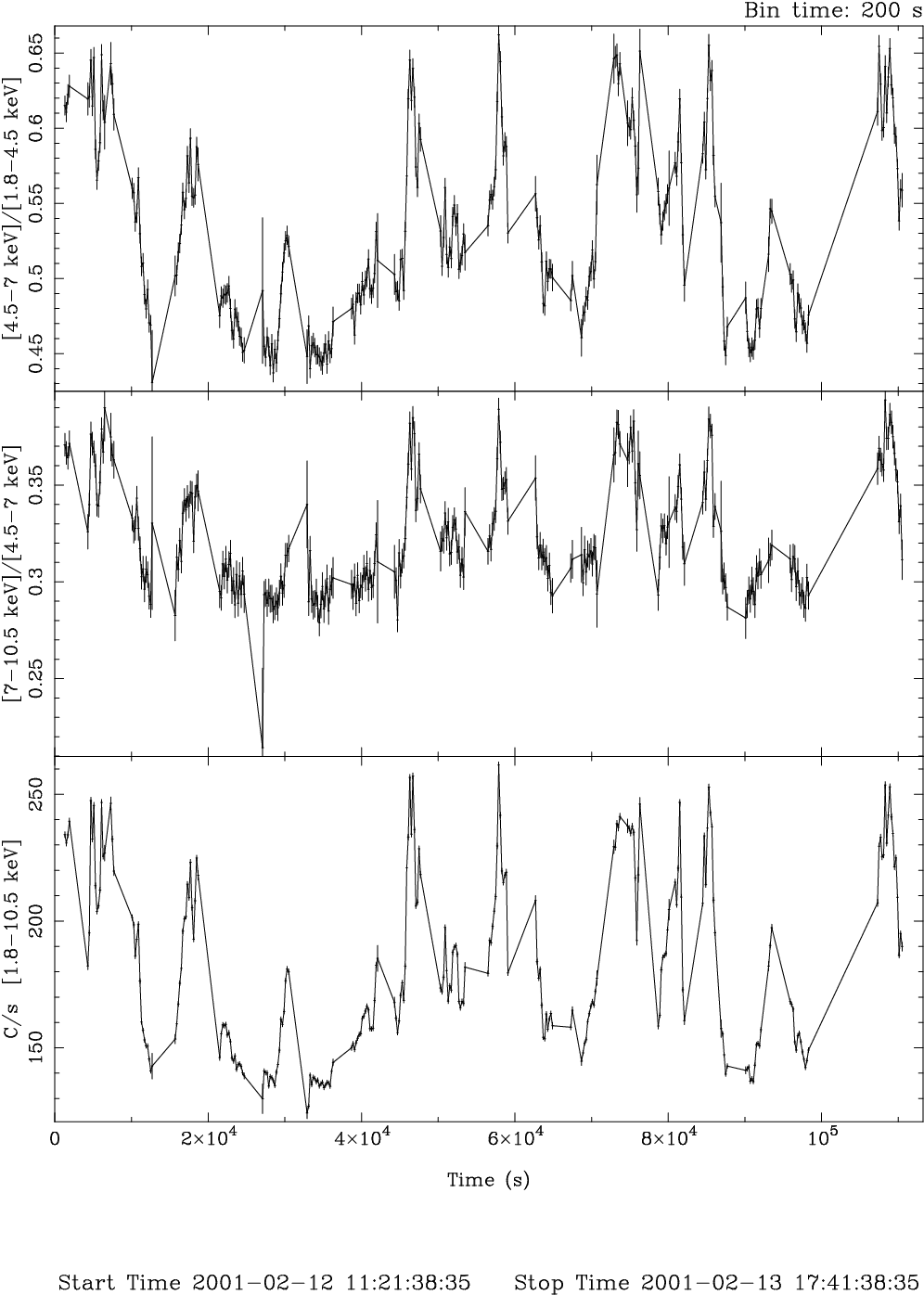}
\caption{Soft color (upper panel), hard color (middle panel), and source
count rate (lower panel) versus time during the first part of the BeppoSAX
observation of GX 349+2.
  \label{fig1}}
\end{figure}

There might be, however, another explanation for the difference in the hard
X-ray component between the March 2000 and the February 2001 BeppoSAX 
observations. From the light curves shown in Figures~\ref{fig1} and 
\ref{fig2} it can be seen that the source was continuously flaring diring
the last BeppoSAX observation. The longest continuous time interval that the
source spent in a persistent flux level (below $\sim 200$ counts/s) is 
$\sim 25$ ks. On the other hand, the source spent a longer continuous time
in a persistent flux level during the previous observation, showing only one
big (i.e.\ with a count rate higher than 200 counts/s) flare during almost 
100 ks of total observation time (see a light curve in Fig.~1 of Paper I). 
It can therefore be that the continuous flaring activity during the last 
BeppoSAX observation has quenched the hard X-ray emission. In other words, 
while the source was mainly in the vertex between the NB and 
the FB (where the hard X-ray component was indeed significantly detected) 
during the previous BeppoSAX, with only a short-lasting excursion in
the FB, the source was mainly in the FB during the last BeppoSAX observation.
In this case the non-detection of the hard component during the last
BeppoSAX observation is in agreement with the non-detection of the
hard component during the flaring state in the previous BeppoSAX observation 
(although the relatively low statistics, due to the short exposure time,
did not allow a definitive conclusion in the previous observation) and with 
the paradigm that the hard emission is suppressed at the highest inferred 
mass accretion rates.

\acknowledgments
This work was partially supported by the Italian Space Agency (ASI) and by 
the Ministero della Istruzione, della Universit\`a e della Ricerca (MIUR). 
TD acknowledges the Netherlands Organization for Scientific Research (NWO).

\clearpage

\begin{deluxetable}{lccccc} 
\tabletypesize{\scriptsize}
\tablecaption{\label{tab:TABexpo} Count rates and exposure times during the 
BeppoSAX observation of GX 349+2.
In column 2 we report the the count rate of GX 349+2 corresponding  to each 
selected region.
The exposure times  are shown in columns  3, 4, 5 and 6 for each of the
BeppoSAX NFIs, respectively. In line 6 we show the total exposure time for 
each NFI. 
} 
\tablewidth{0pt}
\tablehead{\colhead{$~$} 
&\colhead{Counts s$^{-1}$}&\colhead{LECS}&\colhead{MECS}&\colhead{HP}
&\colhead{PDS}\\ \colhead{$~$} 
&\colhead{1.8--10.5 keV}&\colhead{ks}&\colhead{ks}&\colhead{ks}
&\colhead{ks}} 
\startdata 
Interval 1  
& $133.0 \pm 9.0$   &5 &  28 & 26 & 13  \\

Interval 2 
& $152.0 \pm 9.0$ & 5 &  30 & 27 & 14  \\

Interval 3 
& $176 \pm 15$ & 8 &  38 & 35 & 18  \\

Interval 4 
& $206 \pm 15$ & 7 &  22 & 19 & 10  \\

Interval 5 
& $242 \pm 21$ & 5 &  18 & 16 & 8  \\
\tableline
TOT 
& & 30 &  136 & 123 & 63  \\

\tableline 
\enddata 
\end{deluxetable}

\begin{deluxetable}{lc} 
\tabletypesize{\scriptsize}
\tablecaption{\label{tab:TABavera} Best fit parameters of the continuum 
emission of GX 349+2 obtained from the averaged spectrum in the 0.12--200 
keV energy band. 
The continuum consists of a blackbody, a Comptonized spectrum modeled by 
Comptt, and a power law.  kT$_{\rm BB}$ and N$_{\rm BB}$ are, respectively, 
the blackbody temperature and normalization in units of L$_{39}$/D$_{10}$,
where L$_{39}$ is the luminosity in units of $10^{39}$ ergs s$^{-1}$
and D$_{10}$ is the distance in units of 10 kpc.  kT$_0$ is the
temperature of the seed photons for the Comptonization, kT$_e$ is the
electron temperature, $\tau$ is the optical depth of the scattering
cloud using a spherical geometry, N$_{\rm comptt}$ is the normalization of
the Comptt model in XSPEC v.11 units, R$_{\rm W}$ is the Wien radius of the
seed photons in km, and Flux$_{\rm comptt}$ is the bolometric flux 
corresponding to the Comptonization model in units of ergs cm$^{-2}$ s$^{-1}$. 
The power-law normalization is in units of photons keV$^{-1}$ cm$^{-2}$ 
s$^{-1}$ at 1 keV and its flux is indicated as  Flux$_{\rm po}$. 
Uncertainties are at 90\% confidence level for a single parameter of 
interest, the uncertainty on the power-law flux, Flux$_{\rm po}$, is at 
1$\sigma$ confidence level.  }
\tablewidth{0pt}
\tablehead{\colhead{Continuum parameter} 
&\colhead{Value}} 
\startdata 
$N_{\rm H}$ $\rm (\times 10^{22}\;cm^{-2})$ 
& $0.673^{+0.048}_{-0.013}$ \\

kT$_{BB}$ (keV)
& $0.546 \pm 0.016$ \\

N$_{BB}$ $ (\times 10^{-2})$ 
& $5.38^{+0.23}_{-0.25}$  \\

R$_{BB}$ (km)
& $35 \pm 3$ \\

Flux$_{BB}$  $(\times 10^{-8}$ erg cm$^{-2}$ s$^{-1}$)
& $ 0.45\pm 0.02 $ \\

kT$_{0}$ (keV)
&  $1.197^{+0.033}_{-0.038}$ \\

kT$_{e}$ (keV)
& $2.832^{+0.036}_{-0.050}$ \\

$\tau$ 
& $11.06^{+0.40}_{-0.33}$ \\

N$_{comptt}$
& $1.259^{+0.039}_{-0.031}$ \\

Flux$_{comptt}$  $(\times 10^{-8}$ erg cm$^{-2}$ s$^{-1}$)
& $\sim 1.98$ \\

R$_W$ (km)
& $7.6 \pm 0.5$ \\

Photon Index 
& $1.52^{+1.57}_{-0.98}$  \\

Power-law N ($\times 10^{-3}$)
& $1.6^{+20.2}_{-1.6} $  \\

Flux$_{po}$ [10--60 keV] $(\times 10^{-11}$ erg cm$^{-2}$ s$^{-1}$)
& $2.3 \pm 0.8$  \\

Chance Impr. Prob. Power-law 
& $1.9 \times 10^{-7}$ \\

$\chi^2_{\nu}$ (d.o.f.)
& 0.81 (184)   \\

\tableline 
\enddata 
\end{deluxetable}

\begin{deluxetable}{lc} 
\tabletypesize{\scriptsize}
\tablecaption{\label{tab:TABaverabis} Best fit parameters of the
discrete features in GX 349+2 obtained from the averaged spectrum. 
Three gaussian emission lines and an absorption edge are detected.
The centroids, the widths and the normalizations of the
lines are indicated respectively by E, $\sigma$ and I. The labels 1.2,
2.6, and Fe correspond to the  line at 1.2 keV, 2.6 keV  and 6.7 keV,
respectively. Uncertainties are at 90\% confidence level for a single 
parameter of interest.  }
\tablewidth{0pt}
\tablehead{\colhead{Discrete features parameter} 
&\colhead{Value}} 
\startdata 
E$_{edge}$ (keV)
& $9.09^{+0.59}_{-0.52}$  \\

$\tau_{edge}$  $(\times 10^{-2})$
& $2.73^{+1.52}_{-1.36}$ \\

E$_{Fe}$ (keV)
& $6.750^{+0.092}_{-0.095}$  \\

$\sigma_{Fe}$ (keV)
& $0.24 \pm 0.16$  \\

I$_{Fe}$ ($\times 10^{-3}$ photons cm$^{-2}$ s$^{-1}$)
& $7.0^{+2.8}_{-2.1}$ \\

Fe Equivalent Width (eV)
& $39 \pm 13$\\  

Chance Impr. Prob. I$_{Fe}$
& $ 1.6 \times 10^{-9}$\\

E$_{1.2}$ (keV)
&  $1.181 \pm 0.027$ \\

$\sigma_{1.2}$ (keV)
& $< 0.08$ \\

I$_{1.2}$  ($\times 10^{-2}$ photons cm$^{-2}$ s$^{-1}$)
& $1.49^{+0.70}_{-0.36}$\\

Emission--line Equivalent Width (eV)
& $13^{+6}_{-3}  $\\  

Chance Impr. Prob. I$_{1.2}$
& $ 1.1 \times 10^{-8}$\\

E$_{2.6}$ (keV)
&  $2.595 \pm 0.055$ \\

$\sigma_{2.6}$ (keV)
& $< 0.13$ \\

I$_{2.6}$  ($\times 10^{-3}$ photons cm$^{-2}$ s$^{-1}$)
& $6.7^{+3.0}_{-2.4}$\\

Emission--line Equivalent Width (eV)
& $10 \pm 4$\\  

Chance Impr. Prob. I$_{2.6}$
& $ 2.1 \times 10^{-5}$\\

\tableline 
\enddata 
\end{deluxetable} 


\begin{deluxetable}{lccccc} 
\tabletypesize{\scriptsize}
  \tablecaption{\label{tab:TAB2} 
The 0.12--200 keV continuum best fit parameters of GX~349+2 obtained 
from the spectra extracted in different positions in the CD. 
The parameters are defined as in Table \ref{tab:TABavera}. Uncertainties 
are at 90\% confidence level for a single parameter of 
interest except that the uncertainty of  Flux$_{po}$ that is at  1$\sigma$ }  
\tablewidth{0pt}
  \tablehead{\colhead{Continuum parameter} &\colhead{Interval 1}
    &\colhead{Interval 2}&\colhead{Interval 3}&\colhead{Interval 4}
    &\colhead{Interval 5}} 
\startdata

  $N_{\rm H}$ $\rm (\times 10^{22}\;cm^{-2})$ &
  $0.666^{+0.027}_{-0.024}$ & $0.671^{+0.025}_{-0.020}$ &
  $0.664^{+0.016}_{-0.015}$ & $0.686^{+0.038}_{-0.017}$
  & $0.683^{+0.025}_{-0.019}$ \\

kT$_{BB}$ (keV)
& $0.508^{+0.011}_{-0.018}$ 
& $0.526^{+0.011}_{-0.018}$ 
& $0.537 \pm 0.012$ 
& $0.549^{+0.012}_{-0.007}$
& $0.568 \pm 0.013$ \\

N$_{BB}$ $ (\times 10^{-2})$ 
& $4.60^{+0.28}_{-0.21}$ 
& $5.06^{+0.13}_{-0.19}$ 
& $5.21^{+0.15}_{-0.16}$ 
& $5.73 \pm 0.18$
& $5.97 \pm 0.18$  \\

R$_{BB}$ (km)
& $37 \pm 3$ 
& $36 \pm 3$ 
& $36 \pm 2$ 
& $35 \pm 2$ 
& $34 \pm 2$ \\

Flux$_{BB}$  $(\times 10^{-8}$ erg cm$^{-2}$ s$^{-1}$)
& $  0.39 \pm 0.02$ 
& $ 0.42 \pm 0.01$ 
& $ 0.44 \pm 0.01$ 
& $ 0.48 \pm 0.02 $ 
& $ 0.50 \pm 0.02 $ \\

kT$_{0}$ (keV)
&  $1.030^{+0.042}_{-0.039}$
&  $1.119^{+0.021}_{-0.032}$ 
&  $1.176^{+0.024}_{-0.026}$
&  $1.238^{+0.026}_{-0.025}$
&  $1.298 \pm 0.029$ \\

kT$_{e}$ (keV)
& $2.722 \pm 0.042$
& $2.807^{+0.040}_{-0.047}$
& $2.784 \pm 0.034$
& $2.810^{+0.033}_{-0.047}$
& $2.811 \pm 0.042$ \\

$\tau$ 
& $11.24^{+0.36}_{-0.34}$
& $10.58^{+0.36}_{-0.28}$
& $11.04^{+0.30}_{-0.28}$ 
& $11.55^{+0.33}_{-0.31}$
& $12.35 \pm 0.34 $ \\

N$_{comptt}$
& $1.031^{+0.049}_{-0.043}$ 
& $1.085^{+0.044}_{-0.037}$
& $1.271 \pm 0.037$ 
& $1.542^{+0.042}_{-0.037}$
& $1.783^{+0.039}_{-0.040}$ \\

Flux$_{comptt}$  $(\times 10^{-8}$ erg cm$^{-2}$ s$^{-1}$)
& $ \sim 1.45 $ 
& $ \sim 1.64 $ 
& $ \sim 1.94 $ 
& $ \sim 2.48 $ 
& $ \sim 3.00 $ \\

R$_W$ (km)
& $8.9 \pm 0.7$ 
& $8.8 \pm 0.7$ 
& $7.9 \pm 0.4$ 
& $7.8 \pm 0.4$ 
& $7.4 \pm 0.4$ \\

Photon Index 
&  1.52 (fixed)  
&  1.52 (fixed) 
&  1.52 (fixed)
&  1.52 (fixed)
&  1.52 (fixed) \\

Power-law N ($\times 10^{-3}$)
& $2.02 \pm 1.04$  
& $1.21 \pm 1.10$ 
& $2.23^{+0.77}_{-1.08}$ 
& $< 1.27$ 
& $1.83 \pm 1.40$\\

Flux$_{po}$ [10--60 keV] $(\times 10^{-11}$ erg cm$^{-2}$ s$^{-1}$)
& $\sim 2.8$ 
& $\sim 1.7$ 
& $\sim 3.1$ 
& $< 1.4$ 
& $\sim 2.5$ \\

Chance Impr. Prob. Power-law 
& $ 2.5 \times 10^{-3}$ 
& 0.20
& $ 9.4 \times 10^{-4}$ 
& 0.13 
& 0.06\\

$\chi^2_{\nu}$ (d.o.f.)
& 1.06 (184) 
& 1.26 (183) 
& 1.24 (184)
& 1.29 (186) 
& 1.14 (187)  \\

\tableline 
\enddata 
\end{deluxetable} 

\begin{deluxetable}{lccccc} 
\tabletypesize{\scriptsize}
  \tablecaption{\label{tab:TAB2bis} Best fit parameters of 
the discrete features of GX~349+2 obtained from the spectra extracted 
in different positions in the CD. 
The parameters are defined as in Table \ref{tab:TABaverabis}. Uncertainties 
are at 90\% confidence level for a single parameter of 
interest  }  
\tablewidth{0pt}
  \tablehead{\colhead{Discrete features parameter} &\colhead{Interval 1}
    &\colhead{Interval 2}&\colhead{Interval 3}&\colhead{Interval 4}
    &\colhead{Interval 5}} 
\startdata

E$_{edge}$ (keV)
& $9.15 \pm 0.19$ 
& $8.98^{+0.29}_{-0.26}$ 
& $9.05^{+0.42}_{-0.25}$
& $9.53^{+0.20}_{-0.15}$ 
& $9.43  \pm 0.30$  \\

$\tau_{edge}$  $(\times 10^{-2})$
& $4.8 \pm 1.3$ 
& $3.7^{+1.7}_{-1.2}$  
& $3.4 \pm 1.0$
& $3.9^{+1.3}_{-1.2}$ 
& $3.8 \pm 1.1$ \\

E$_{Fe}$ (keV)
& $6.710^{+0.055}_{-0.060}$ 
& $6.735 \pm 0.039$
& $6.746 \pm 0.053$ 
& $6.765 \pm 0.068$
& $6.711 \pm 0.094$ \\

$\sigma_{Fe}$ (keV)
& $0.39^{+0.10}_{-0.09}$ 
& $0.208 \pm 0.075$
& $0.244^{+0.082}_{-0.073}$ 
& $< 0.30$ 
& $< 0.29$ \\

I$_{Fe}$ ($\times 10^{-3}$ photons cm$^{-2}$ s$^{-1}$)
& $9.2^{+2.2}_{-1.6}$ 
& $7.6^{+1.4}_{-1.3}$
& $7.2^{+1.6}_{-1.3}$  
& $6.3^{+1.8}_{-1.5}$
& $4.9^{+1.9}_{-1.7}$ \\

Fe Equivalent Width (eV)
& $77 \pm 15$ 
& $52 \pm 11$ 
& $40 \pm 8$
& $28 \pm 8$ 
& $18 \pm 7$\\  

Chance Impr. Prob. I$_{Fe}$
& $\sim 0$ 
& $\sim 0$ 
& $\sim 0 $ 
& $ 2.2 \times 10^{-10}$
& $ 1.2 \times 10^{-5}$\\

E$_{1.2}$ (keV)
&  $1.180$ (fixed) 
&  $1.181 \pm 0.027$
&  $1.190^{+0.030}_{-0.036}$ 
&  $1.15^{+0.10}_{-0.22}$  
&  $1.21^{+0.06}_{-0.14}$\\

$\sigma_{1.2}$ (keV)
& $< 0.23$ 
& $< 0.08$ 
& $< 0.08$ 
& $< 0.29$
& $< 0.20$ \\

I$_{1.2}$  ($\times 10^{-2}$ photons cm$^{-2}$ s$^{-1}$)
& $2.1^{+4.3}_{-0.9}$
& $2.6^{+1.0}_{-0.8}$ 
& $1.42^{+0.70}_{-0.49}$ 
& $1.1^{+4.8}_{-0.6}$ 
& $1.1^{+1.3}_{-0.7}$\\

Emission--line Equivalent Width (eV)
& $20^{+38}_{-9} $ 
& $23 \pm 8 $
& $13 \pm 6 $
& $9^{+40}_{-5}  $
& $9^{+11}_{-6}  $\\  

Chance Impr. Prob. I$_{1.2}$
& $ 2.0 \times 10^{-4}$
& $ 9.4 \times 10^{-7}$
& $ 5.0 \times 10^{-4}$ 
& 0.11
& 0.11\\

E$_{2.6}$ (keV)
&  $2.629 \pm 0.070$ 
&  $2.613^{+0.038}_{-0.059}$
&  $2.622^{+0.059}_{-0.064}$ 
&  $2.636^{+0.050}_{-0.056}$  
&  $2.620^{+0.070}_{-0.074}$\\

$\sigma_{2.6}$ (keV)
& $< 0.32$ 
& $< 0.12$ 
& $< 0.12$ 
& $< 0.10$
& $< 0.13$ \\

I$_{2.6}$  ($\times 10^{-3}$ photons cm$^{-2}$ s$^{-1}$)
& $4.7^{+3.0}_{-2.1}$
& $4.6^{+2.7}_{-0.9}$ 
& $5.2^{+2.3}_{-1.9}$ 
& $6.0^{+2.4}_{-2.0}$ 
& $4.8^{+2.6}_{-2.4}$\\

Emission--line Equivalent Width (eV)
& $8 \pm 4$ 
& $7^{+4}_{-1} $
& $7^{+4}_{-2} $
& $8 \pm 3$
& $6 \pm 3$\\  

Chance Impr. Prob. I$_{2.6}$
& $ 7.9 \times 10^{-3}$
& $ 4.5 \times 10^{-3}$
& $ 2.0 \times 10^{-3}$ 
& $ 1.8 \times 10^{-3}$
& 0.04\\

\tableline 
\enddata 
\end{deluxetable} 

\begin{deluxetable}{lccccc} 
\tabletypesize{\scriptsize}
\tablecaption{\label{tab:line} 
Radius in cm and electron density in cm$^{-3}$ of the region in which the 
emission lines are produced.  }
\tablewidth{0pt}
  \tablehead{\colhead{line} &\colhead{Interval 1}&\colhead{Interval 2}
    &\colhead{Interval 3}&\colhead{Interval 4}&\colhead{Interval 5}     } 
\startdata 

S XVI & $r = 8.6 \times 10^9$    &  $r = 1.1 \times 10^{10}$ & 
        $r = 1.3 \times 10^{10}$ &  $r = 1.7 \times 10^{10}$ & 
        - \\
      & $n_e = 5.9 \times 10^{14}$ & $n_e = 4.4 \times 10^{14}$ &
        $n_e =3.5 \times 10^{14}$ & $n_e = 2.5 \times 10^{14}$ &
        - \\

\tableline

Fe XXIV &  $r = 1.1 \times 10^{10}$  &  $r = 1.1 \times 10^{10}$ 
        &  $r = 2.7 \times 10^{10}$  & - & - \\
        &  $n_e = 4.4 \times 10^{14}$ &  $n_e = 4.9 \times 10^{14}$ 
        &  $n_e = 9.0 \times 10^{13}$ & - & - \\

\tableline 

Fe XXV &  $r = 2.3 \times 10^9$    &  $r = 3.4 \times 10^{9}$ &
          $r = 4.8 \times 10^9$    &  $r = 9.5 \times 10^{9}$ &        
          $r = 1.5 \times 10^{10}$ \\
       &  $n_e = 2.2 \times 10^{15}$ &  $n_e = 1.2 \times 10^{15}$&
          $n_e = 6.5 \times 10^{14}$ &  $n_e = 2.6 \times 10^{14}$&
          $n_e = 9.8 \times 10^{13}$ \\ 
       
\tableline 
\enddata 
\end{deluxetable}

\clearpage
 
\begin{figure}
\plotone{f2.ps}
\caption{Soft color (upper panel), hard color (middle panel), and source
count rate (lower panel) versus time during the second part of the BeppoSAX
observation of GX 349+2. 
  \label{fig2}}
\end{figure}

\begin{figure}
\plotone{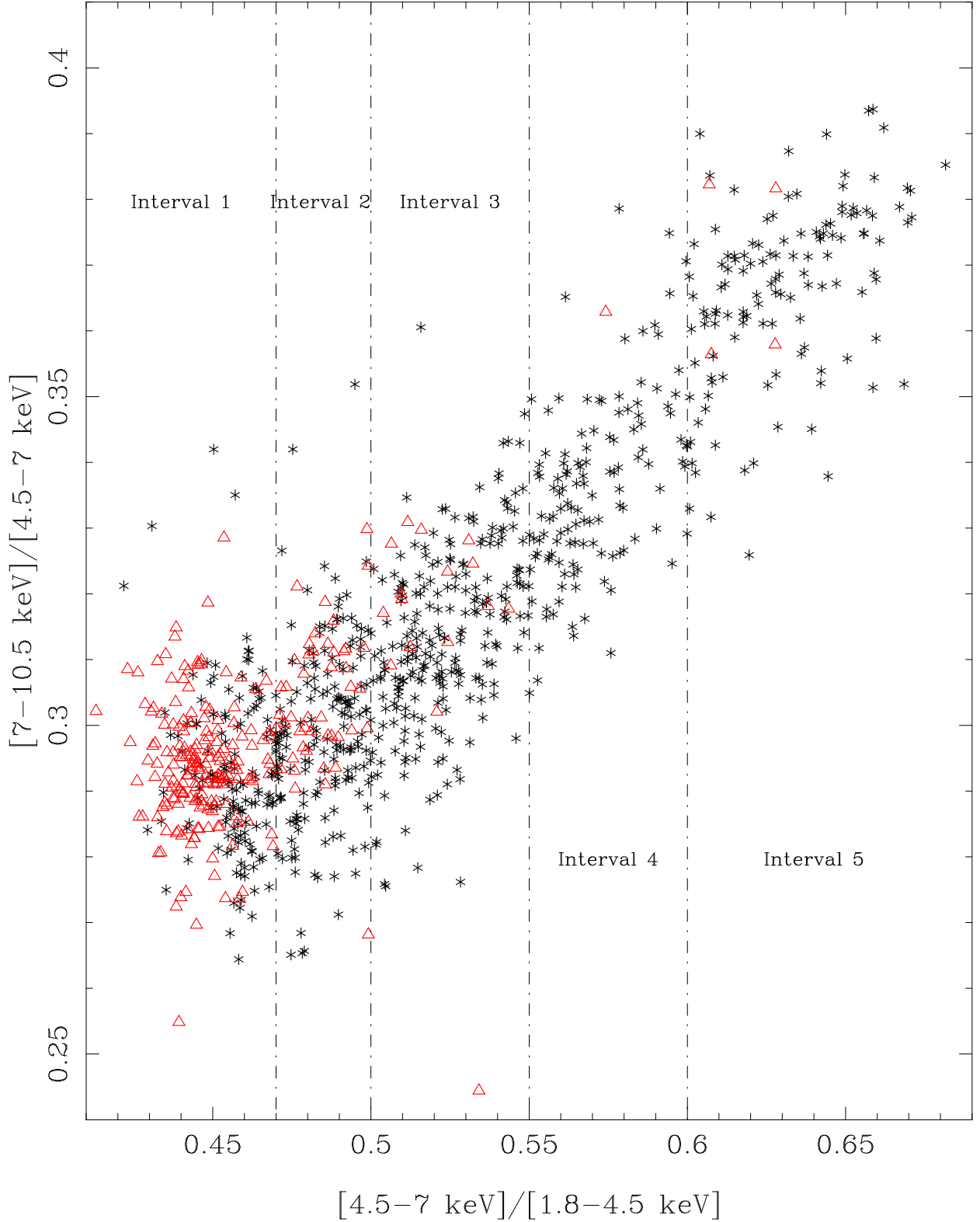}
\caption{  Color-Color Diagram of GX~349+2. Each bin corresponds
to 200 s.  The red triangles and the black stars indicate the position
of the source obtained from the observation showed in Paper I and from
our observations, respectively. The intervals, in which
the CD is divided, are indicated by the dot-dashed vertical lines.
\label{fig3zero}}
\end{figure}

\begin{figure}
\plotone{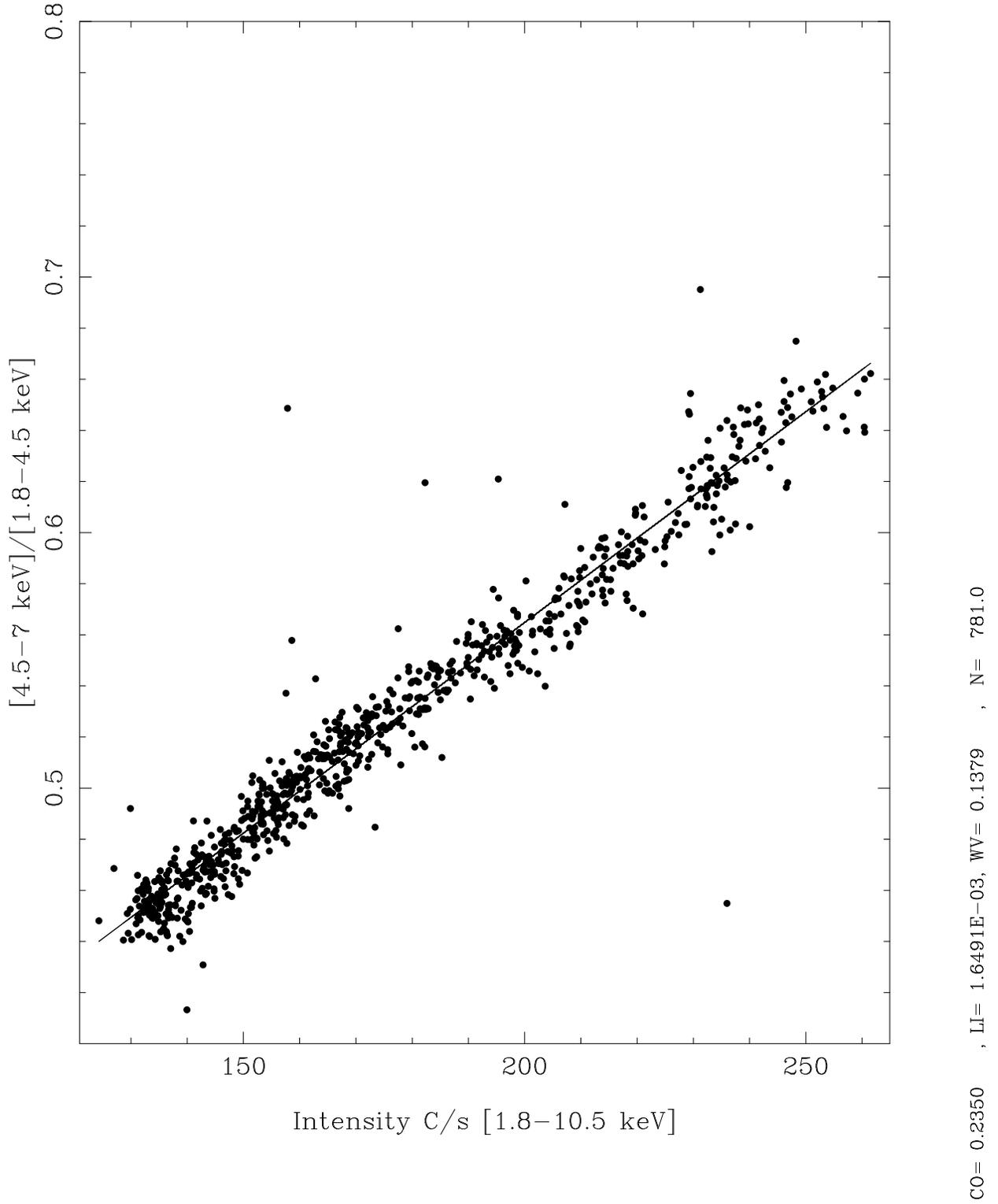}
\caption{ Hardness-Intensity  Diagram of GX~349+2.
  Each bin corresponds to 200 s.  The solid line indicates the best
 fit to the data using a linear relation $y= 1.65 \times 10^{-3} x +
 0.235$. \label{fig3bis}}
\end{figure}

\begin{figure}
\plotone{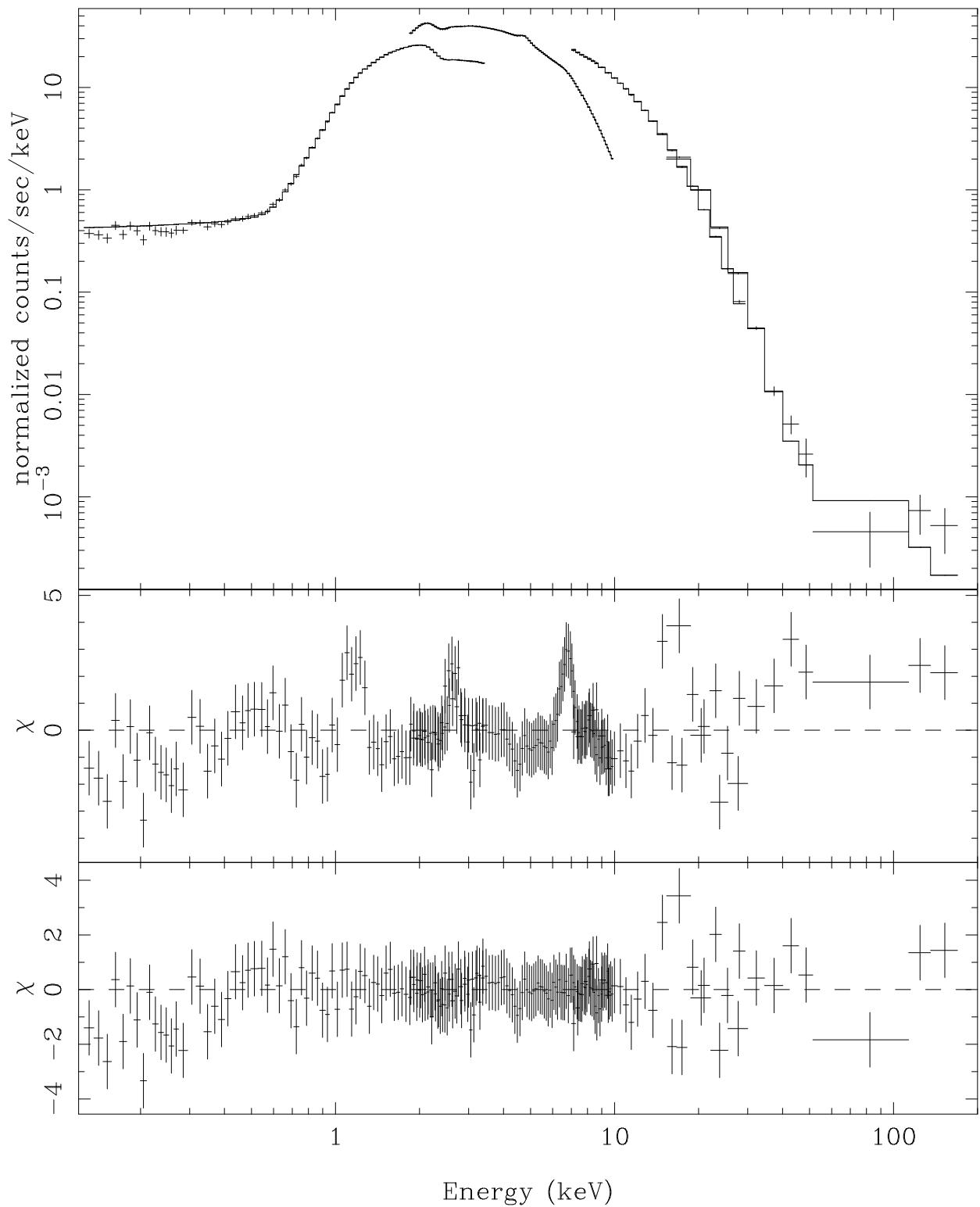}
\caption{{\bf Upper Panel:} Averaged spectrum of GX~349+2 shown together
with the best-fit model. {\bf Middle Panel:} Residuals in unit of $\sigma$ 
with respect to a simple model consisting of blackbody plus Comptt;
three  narrow features at 1.2 keV, 2.6 keV and 6.7 keV, an absorption  
feature at around 9 keV  and a hard excess 
above 30 keV are present.  {\bf Lower Panel:} Residuals in unit of $\sigma$ 
with respect to the best-fit model. \label{fig4}}
\end{figure}

\begin{figure}
\plotone{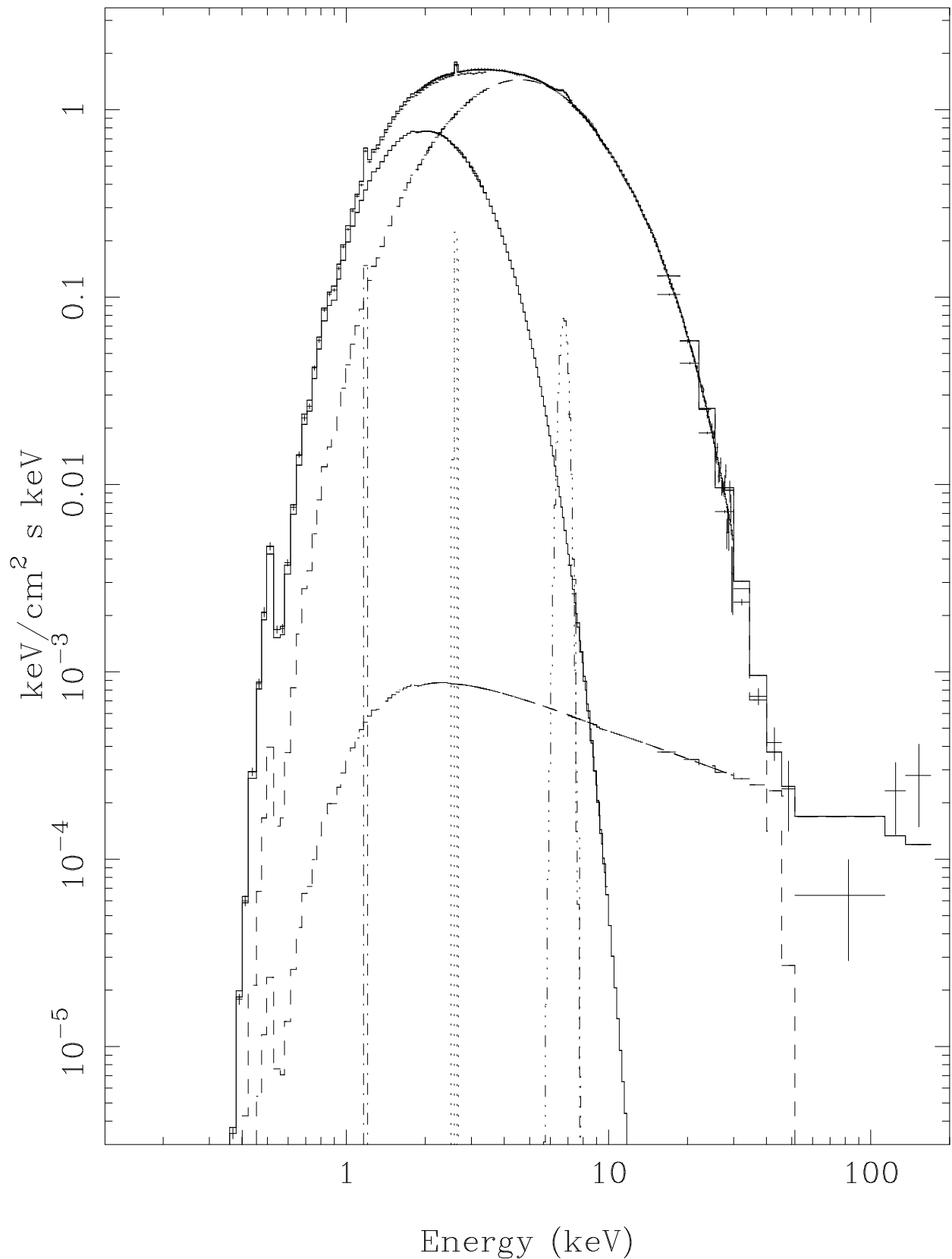}
\caption{Unfolded averaged spectrum of GX~349+2, and the best fit model 
indicated  by the solid line  
on top of the data. The individual model components are also 
shown, namely the blackbody (solid line), the Comptonized spectrum 
({\tt Comptt} model, dashed line), three gaussian emission lines at 
$\sim 1.2$ keV, $\sim 2.6$ keV and $\sim 6.7$ keV, and
the hard power-law (dashed line). \label{fig6}}
\end{figure}

\begin{figure}
\plotone{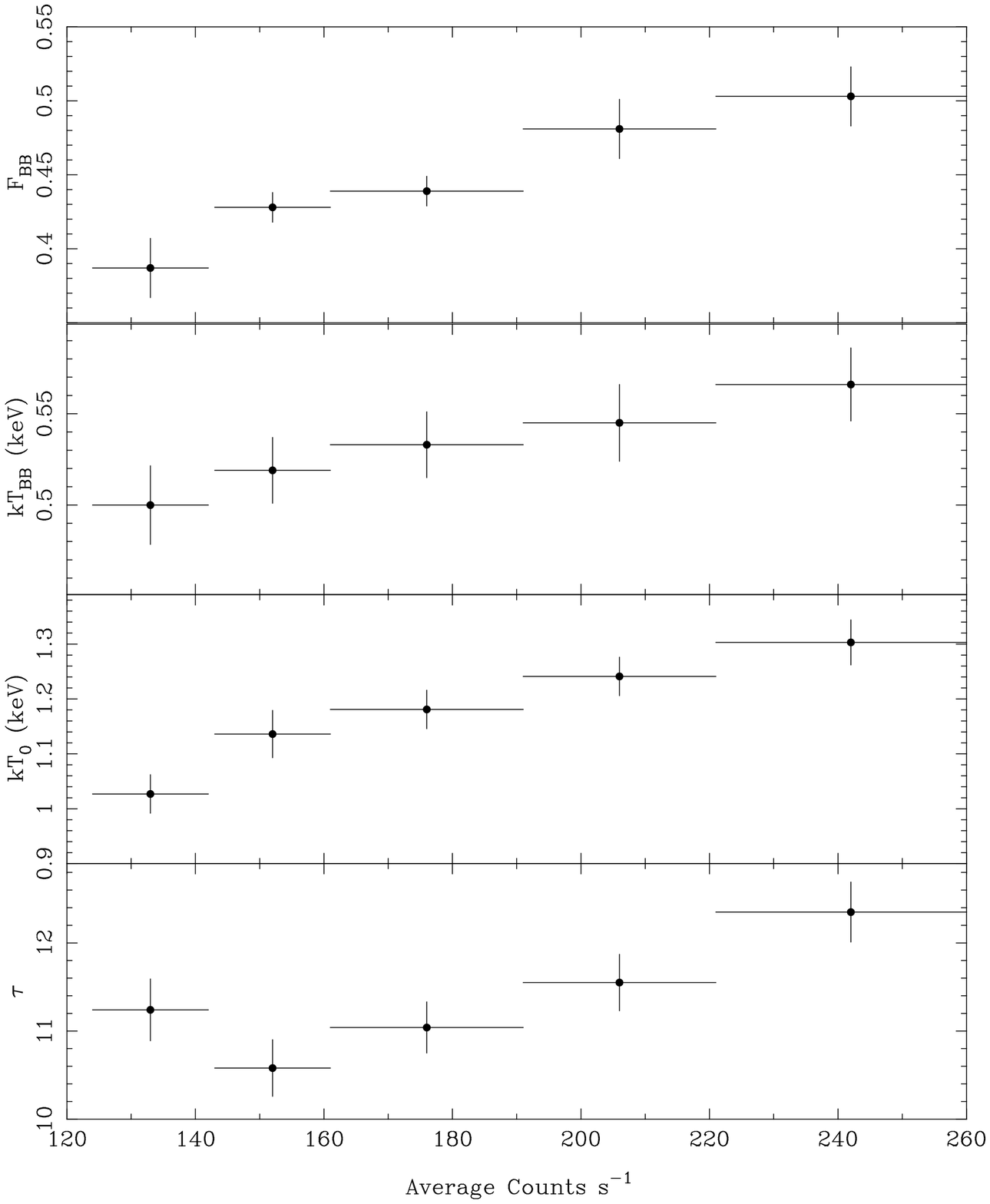}
\caption{ Evolution of the spectral parameters as a function of the
source luminosity/position in the CD.
From the top to the bottom: blackbody flux in units of 
$10^{-8}$ erg cm$^{-2}$ s$^{-1}$,  blackbody temperature in keV, 
seed-photon
temperature in keV and optical depth of the Comptonizing cloud.  
 \label{trend}}
\end{figure}


\clearpage

\begin{thebibliography}{}
\bibitem[]{}
Agrawal, P. C., et al., 1971, Ap\&SS, 10, 500
\bibitem[]{}
Asai, K., et al., 1994, PASJ, 46, 479
\bibitem[]{}
Barret, D., \& Vedrenne, G., 1994, ApJS, 92, 505
\bibitem{bar02} 
Barret, D.,  \&  Olive, J. F.,  2002,  ApJ, 576, 391
\bibitem[]{}
Boella, G., Butler, R. C., Perola, G. C., Piro, L., Scarsi, L., Blecker, J.,
1997a, A\&AS, 122, 299
\bibitem[]{}
Boella, G., et al., 1997b, A\&AS, 122, 327
\bibitem[]{}
Chen, X. 1995, ApJ, 448, 803
\bibitem[]{}
Christian, D. J., \& Swank, J. H., 1997, ApJS, 109, 177
\bibitem[]{}
Cooke, B. A., Ponman, T. J., 1991, A\&A, 244, 358
\bibitem[]{}
D'Amico, F., Heindl, W. A., Rothschild, R. E., Gruber, D. E., 2001, ApJL, 
547, L147
\bibitem[]{}
Di Salvo, T., et al., 2000, ApJL, 544, L119
\bibitem[]{}
Di Salvo, T., et al., 2001, ApJ, 544, 49
\bibitem{dis02} 
 Di Salvo, T.,  Farinelli, R.,  Burderi, L., et al., 2002, A\&A, 386, 535
\bibitem[]{} 
 Di Salvo, T., \& Stella, L., 2002, proceedings of the XXII Moriond
 Astrophysics Meeting "The Gamma-Ray Universe" (Les Arcs, March 9-16,
 2002), eds. A. Goldwurm, D. Neumann, and J. Tran Thanh Van, The Gioi
 Publishers (Vietnam), (astro-ph/0207219)
\bibitem{for96} 
 Ford, E. C.,  Kaaret, P.,  Tavani, M., et al., 1996, ApJ, 469, L37
\bibitem[]{}
Frontera, F., et al., 1997, A\&AS, 122, 357
\bibitem[]{}
Frontera, F., et al., 1998, Nuclear Physics B (Proc. Suppl.), 69, 286
\bibitem[]{}
Greenhill, J. G., Coe, M. J., Burnell, S. J. B., Strong, K. T., 
Carpenter, G. F., 1979, MNRAS, 189, 563
\bibitem[]{}
Hakkila, J., et al., 1997, AJ, 114, 2043 
\bibitem[]{}
Hasinger, G., \& van der Klis, M., 1989, A\&A, 225, 79
\bibitem[]{}
Hasinger, G., van der Klis, M., Ebisawa, K., Dotani, T., Mitsuda, K.,
1990, A\&A, 235, 131
\bibitem[]{}
Haymes, R. C., Harnden, F. R., Johnson, W. N., Prichard, H. M., Bosch, H. E.,
1972, ApJ, 172, L47
\bibitem{hom} 
 Homan, J.,   Wijnands, R.,  van der Klis, M., et al., 2001,
ApJS, 132, 377
\bibitem[]{}
Iaria, R., Burderi, L., Di Salvo, T., La Barbera, A., Robba, N. R., 2001,
ApJ, 547, 412 
\bibitem[]{}
In't Zand, J. J. M., et al., 1999, A\&A, 345, 100
\bibitem[]{}
Jain, A., et al., 1984, A\&A, 140, 179
\bibitem[]{}
Kallman, T. R., Liedahl, D., Osterheld, A., Goldstein, W., Kahn, S., 
1996, ApJ, 465, 994
\bibitem[]{}
Krolik, J. H., McKee, C. F., \& Tarter, C. B., 1981, ApJ, 249, 422  
\bibitem[]{}
Kuulkers, E., et al., 1994, A\&A, 289, 795 
\bibitem[]{}
Kuulkers, E., \& van der Klis, M., 1995, A\&A, 303,801 
\bibitem[]{}
Manzo, G., Giarrusso, S., Santangelo, A., Ciralli, F., Fazio, G., Piraino, 
S., Segreto, A., 1997, A\&AS, 122, 341
\bibitem[]{}
Miyamoto, S., \& Matsuoka, M., 1977, SSRv, 20, 687
\bibitem[]{}
O'Neill, P. M., Kuulkers, E., Sood, R. K., van der Klis, M. 2002,
MNRAS, 336, 217
\bibitem[]{}
O'Neill, P. M., Kuulkers, E., Sood, R. K., Dotani, T. 2001,
A\&A, 370, 479
\bibitem[]{}
Parmar, A. N., et al., 1997, A\&AS, 122, 309
\bibitem[]{}
Penninx, W., Augusteijn, T., 1991,  A\&A, 246, L81 
\bibitem{pet73}
 Peterson, L. E.,  1973, IAU Symp., 55, 51
\bibitem[]{}
Peterson, L. E., \& Jacobson, A. S., 1966, ApJ, 145, 962
\bibitem[]{}
Piraino, S., Santangelo, A., Ford, E. C., Kaaret, P., 1999, A\&A, 349, L77
\bibitem[]{}
Predehl, P., \& Schmitt, J. H.M. M., 1995, A\&A, 293, 889
\bibitem[]{}
Riegler, G. R., Boldt, E., \& Serlemitsos, P., 1970, Nature, 266, 1041 
\bibitem[]{}
Soong, Y., Rothschild, R. E., 1983, ApJ, 274, 327
\bibitem[]{}
Strickman, M., \& Barret, D., Detections of multiple hard X-ray 
flares from Sco X-1 with OSSE, in AIP Conf. Proc. 510, Proc.  of the 
Fifth Compton Symposium, Eds M.L. McConnel and J.M. Ryan 
(New York:AIP), 222-226, 2000
\bibitem[]{}
Tanaka, Y., 1989, The 23rd ESLAB Symposium on Two Topics in X Ray
Astronomy. Volume 1: X Ray Binaries p 3-13
\bibitem[]{}
Titarchuk, L., 1994, ApJ, 434, 570
\bibitem[]{}
Ubertini, P., Bazzano, A., Cocchi, M., La Padula, C., Sood, R. K.,
1992, ApJ, 386, 710
\bibitem[]{}
Valinia, A., \& Marshall, F. E., 1998, ApJ, 505, 134 
\bibitem{van95} 
 van der Klis, M., in: X-Ray Binaries, Cambridge University Press, 
252, 1995.
\bibitem{van00} 
 van der Klis, M., 2000, ARA\&A, 38, 717
\bibitem{van01} 
 van der Klis, M., 2001, ApJ, 561, 943
\bibitem[]{}
Verner, D. A., \& Ferland, G. J., 1996, ApJS, 103, 467 
\bibitem[]{}
White, N. E., \& Holt, S. S., 1982, ApJ, 257, 318
\bibitem[]{}
Zhang, W., Strohmayer, T. E., Swank, J. H., 1998, ApJ, 500, L167
\end{thebibliography}
\end{document}